\RequirePackage{fix-cm}
\documentclass[smallextended]{svjour3} 
\smartqed 
\usepackage{times}
\usepackage{a4}
\usepackage{verbatim}
\usepackage{graphicx}
\usepackage{amssymb,amsmath}
\usepackage{xspace,hyperref}
\usepackage[affil-it]{authblk}

\usepackage{natbib}
\bibpunct{(}{)}{;}{a}{,}{,}
\newcommand{\ro}[1]{\ensuremath{\textrm{#1}}}
\newcommand{\ten}[1]{\ensuremath{\times 10^{#1}}}

\newcommand{\dd}{\ensuremath{\ \textrm{d}}\xspace}
\newcommand{\aT}{\ensuremath{\boldsymbol{\alpha}_T}\xspace}
\newcommand{\bT}{\ensuremath{\boldsymbol{\beta}_T}\xspace}

\title{Fine-Tuning in the Context of Bayesian Theory Testing}

\author{Luke A. Barnes}

\institute{L. Barnes \at
 Sydney Institute for Astronomy \\
School of Physics, University of Sydney \\
NSW 2006, Australia
Tel.: +61 2 90367980
\email{luke.barnes@sydney.edu.au}
}

\date{Received: date / Accepted: date}

\begin{document}
\maketitle

\begin{abstract}
Fine-tuning in physics and cosmology is often used as evidence that a theory is incomplete. For example, the parameters of the standard model of particle physics are ``unnaturally'' small (in various technical senses), which has driven much of the search for physics beyond the standard model. Of particular interest is the fine-tuning of the universe for life, which suggests that our universe's ability to create physical life forms is improbable and in need of explanation, perhaps by a multiverse. This claim has been challenged on the grounds that the relevant probability measure cannot be justified because it cannot be normalized, and so small probabilities cannot be inferred. We show how fine-tuning can be formulated within the context of Bayesian theory testing (or \emph{model selection}) in the physical sciences. The normalizability problem is seen to be a general problem for testing any theory with free parameters, and not a unique problem for fine-tuning. Physical theories in fact avoid such problems in one of two ways. Dimensional parameters are bounded by the Planck scale, avoiding troublesome infinities, and we are not compelled to assume that dimensionless parameters are distributed uniformly, which avoids non-normalizability.

\keywords{Probability \and Bayesian \and Fine-Tuning}
\end{abstract}

\section{Introduction}

Beginning in the 1970's, physicists have calculated that seemingly small changes to the fundamental constants of nature and the initial conditions of the cosmos would have dramatic effects on the universe. In particular, the complexity and stability required by any known or thus-far conceived form of life can be rather easily erased. For example, the masses of the fundamental constituents of ordinary matter --- up quarks, down quarks and electrons --- must be constrained to lie in a very small section of parameter space for nuclei, atoms, molecules, and chemistry to be possible at all. Similarly, the vacuum energy of the universe must be extraordinarily small compared to its ``natural" Planck-scale value for the universe to have any structure. This \emph{fine-tuning of the universe for life} was first investigated by \citet{Carter74}, \citet{Silk1977}, \citet{Carr79}, and \citet{BT86}, and has been reviewed recently by \citet{Hogan2000}, \citet{Barnes2012}, \citet{Schellekens2013} and \citet{Lewis2016}.

The fine-tuning of the universe for life has been pressed into the service of a variety of conclusions --- physical, cosmological and philosophical. For example, fine-tuning has been offered as evidence for a multiverse: the universe as a whole consists of an enormous number of variegated ``pocket" universes, each with different constants and cosmic conditions, and within which observers must see local conditions that are conducive to life forms \citep[see, for example, the brief history given by][]{Linde2015}. It has also been argued that fine-tuning is evidence for a cosmic designer, whose purposes for this universe include the existence of embodied moral agents \citep{Swinburne2004,Collins2009}.

A crucial motivator for these arguments is the intuition that fine-tuning demonstrates that a life-permitting universe is extraordinarily \emph{improbable}. A universe drawn blindly from a big barrel of possible universes is unlikely to have the right forces, particles and cosmic initial conditions for life to develop, it seems. However, a number of philosophers have cast doubt on whether this intuition can be made rigorous. \citet[][hereafter MMV]{MMV} and \citet[][hereafter CGP]{CGP} have argued that the relevant probability measure, because it is spread evenly over an infinitely large range, cannot be normalized and hence the relevant probabilities cannot be calculated. These papers are mostly concerned with the fine-tuning argument for the existence of God, but apply equally to the inference from fine-tuning to the existence of a multiverse. In a similar vein, \citet{Halvorson2014} has argued that a correct understanding of the probabilities shows that a life-permitting universe is unlikely on any assumptions about its origin, and thus fine-tuning cannot be used to argue for anything deeper than the laws of nature. 

Our goal here is to use Bayesian probability theory, as it is employed in theory testing (or \emph{model selection}) in the physical sciences, to show that we can make rigorous the claim that the fine-tuning of the universe for intelligent life renders our universe exceedingly improbable. There really is something to be explained. Precisely what (or Who) that explanation is, is left as an exercise for the reader.

\section{Probabilities for Model Selection in Physics} \label{S:modelselect}
We begin with an overview of model selection in the physical sciences; the reader should be mindful of differences between the approaches to probability in philosophy and physics.

In recent decades, Bayesian approaches to probability theory have significantly changed both the principles and the practice of how physicists analyze data and draw scientific conclusions. Bayesian probabilities $p(B|A)$ in physics are not taken to quantify some aspect of the psychological state of someone who believes $A$ and is considering $B$. Rather, they are presented \citep[for example, by][]{Jaynes2003} as an extension to classical logic, quantifing the \emph{degree of plausibility} of the proposition $B$ given the truth of the proposition $A$. Just as symbolic logic's material condition $A \rightarrow B$ says nothing about whether $A$ is known by anyone, but instead denotes a connection between the truth values of the propositions $A$ and $B$, so $p(B|A)$ quantifies a relationship between these propositions. Note that these are not degrees of truth; $A$ and $B$ are in fact either true or false.

Why think that degrees of plausibility can be modelled by probabilities? There are a number of approaches that lead Bayesians to the probability axioms of \citet{Kolmogorov1933} or similar, such as Dutch book arguments and representational theorems which trace back to \citet{Ramsey1926}. More common among physicists is the theorem of \citet{Cox1946} \citep[see also][]{Jaynes2003,Caticha2009,Knuth2012}. We propose that degrees of plausibility obey the following desiderata :
\begin{itemize} \setlength{\itemsep}{-2pt}
\item[D1.] Degrees of plausibility are represented by real numbers. This ensures that they can be compared on a single scale.
\item[D2.] Degrees of plausibility change in common sense ways. For example, if learning $C$ makes $B$ more likely, but doesn't change how likely $A$ is, then learning $C$ should make ($A$ and $B$) more likely.
\item[D3.] If a conclusion can be reasoned out in more than one way, then every possible way must lead to the same result.
\item[D4.] Information must not be arbitrarily ignored. All given evidence must be taken into account.
\item[D5.] Identical states of knowledge (except perhaps for the labeling of the propositions) should result in identical assigned degrees of plausibility.
\end{itemize}

Cox's theorem shows that degrees of plausibility are probabilities, or more precisely, they obey the usual rules of probability. (Hereafter, we will use the term probabilty unless further distinction is required.) We have a rule for each of the Boolean operations `and' ($AB$), `or' ($A+B$) and 'not' ($\bar{A}$)\footnote{As our presentation of Bayesian probability is somewhat different to the usual philosophical presentation, we've used the notation most familiar to physicists.},
\begin{align}
p(AB|C) &\equiv p(A|BC) ~p(B|C) \equiv p(B|AC) ~p(A|C) \label{eq:prod}\\
p(A+B|C) &\equiv p(A|C) + p(B|C) - p(AB|C) \label{eq:sum} \\
p(\bar{A}|C) &\equiv 1 - p(A|C) ~. \label{eq:not}
\end{align}
From Equation \eqref{eq:prod} we can derive Bayes' theorem (assuming $p(B | C) \neq 0$),
\begin{equation} \label{eq:bayes}
p(A|BC) = \frac{p(B|AC) ~ p(A | C)}{p(B | C)} ~.
\end{equation}
These are \emph{identities}, holding for any propositions $A$, $B$ and $C$ for which the relevant quantities are defined. While Bayes' theorem often comes attached to a narrative about prior beliefs, updating and conditioning, none of this is essential. In fact, an insistence that Bayes' theorem must be applied in chronological order (``updating") is contrary to D3, which is so crucial to the Bayesian (and Coxian) approach that \citet{Skilling2014} goes so far as to claim that ``probability calculus is forced upon us as the only method which lets us learn from data irrespective of their order". Even if one does not agree with Skilling, assigning known propositions to $B$ and $C$ in Equation \eqref{eq:bayes} is purely for convenience.

When Bayes' theorem is used to calculate the probability of some hypothesis or theory $T$, given evidence $E$ and background information $B$, the corresponding terms in Equation \eqref{eq:bayes} are commonly given the following names: $p(T|EB)$ is the \emph{posterior probability}, $p(T|B)$ is the \emph{prior probability}, $p(E|TB)$ is the \emph{likelihood}, and $p(E|B)$ is the \emph{marginal likelihood}. As noted above, the conjunction $EB$ represents everything that the posterior treats as ``known'', and the separation into $E$ and $B$ (into evidence and background) is purely for convenience.

While not our primary focus here, the assignment of prior probabilities is (at best) an active research problem for the Bayesian. Here, we note \emph{one} important aspect of prior probabilities: they are crucial for penalizing ad-hoc theories. Consider a simple case: suppose that a physical theory attempts to ``cheat the likelihood" by simply adding the data to the theory, $T_\textrm{new} = TU$. This gives the new theory a perfect likelihood: $p(U|T_\textrm{new}B) = p(U|TUB) = 1$. While the likelihood is fooled, the posterior is not because it depends on the prior: $p(T_\textrm{new}|B) = p(TU|B) = p(U|TB) p(T|B)$, and thus $p(T_\textrm{new}|UB) = p(T|UB)$. The lesson is this: don't smuggle data into your theory.\footnote{Consistent application of D3 avoids the ``problem of old evidence" for the objective Bayesian. \citet{Glymour1980} argues that, if we already know evidence $E$, then $p(E|T) = 1$ and $p(E) = 1$, and thus $p(T|E) = p(T)$. This is not how to use Bayes theorem. Even if $E$ is known, we should not take $E$ as given in every probability we calculate. The posterior $p(T|EB)$ takes $E$ as given, but the likelihood $p(E|TB)$ and marginal likelihood $p(E|B)$ do not. Calculating the likelihood uses the same probability function that comes from Cox's theorem; it does not require a new ``ur-probability'' function, generated by supposing ``that one does not fully believe that E'' \citep{Monton2006}. Objective Bayesian probabilities aren't about what any individual knows or believes; they are about what one proposition implies about the plausibility of another. Similarly, there is no need to argue in chronological order, taking life as background information and fine-tuning as new information \citep{Roberts2011}.}

We turn now to testing physical theories. Let,
\begin{itemize} \setlength{\itemsep}{-2pt} \renewcommand{\labelitemi}{$\bullet$}
\item $T$= the theory to be tested. For our purposes, the important thing about a physical theory is that it implies certain expectations about physical scenarios. As a specific example, $T$ may represent a set of symmetry principles, from which we can derive the mathematical form of a Lagrangian (or, equivalently, the dynamical equations), but not the values of its free parameters.
\item $U$ = our observations of this Universe.
\item $B$ = everything else we know. For example, the findings of mathematics and theoretical physics are included in $B$. As I have defined it for our purposes here, the information in $B$ does not give us any information about which possible world is actual. The theoretical physicist can explore models of the universe mathematically, without concern for whether they describe reality.
\end{itemize}
To help in calculating the posterior, we turn to Bayes Theorem,
\begin{equation} \label{eq:bayesTUB}
p(T|UB) = \frac{p(U|TB) ~ p(T | B)} {p(U | B)} ~.
\end{equation}

A physical theory describes the physical world, and so should make claims about which physical scenarios (including empirical data) are more or less to be expected. In the Bayesian framework, this implies that calculating \emph{likelihoods} $p(U|TB)$ is part of the job description of any physical theory. Note that there are several, conceptually distinct sources of the uncertainty quantified by the likelihood. Firstly, there may be stochasticity within the theory itself. This could be because the theory is intrinsically indeterminate, such as (some interpretations of) quantum mechanics, or because $T$ is an \emph{effective} theory that describes physical systems by averaging over --- and thus blurring out --- microphysical details. Secondly, our observations are imprecise, and thus consistent with a range of physically possible universes.

We require two further ingredients to calculate $p(U|TB)$. Firstly, physical theories typically contain free parameters, that is, mathematical constants that appear in the Lagrangian. We will call them, collectively, \aT. If the Lagrangian in question is a fundamental theory of physics, that is, a theory that is not an approximation to another known theory, then these parameters are called \emph{fundamental constants of nature}. Secondly, a description of the physical universe is given by a \emph{solution} to the equations of the theory. The set of solutions to $T$ describes the set of universes that are physically possible according to $T$. To specify a particular solution, we usually require a few more numbers (denoted \bT) in the form of initial conditions. (It is a useful simplification to consider \bT to be simply a set of numbers.)

To calculate $p(U|TB)$, we \emph{marginalize} over the constants \aT and \bT, treating them as nuisance parameters. By the law of total probability,
\begin{equation} \label{eq:totalprob}
p(U|TB) = \int p(U | \aT \bT T B) ~ p( \aT \bT | T B) \dd \aT \dd \bT ~.
\end{equation}
We can call $p( \aT \bT | T B)$ the \emph{prior probability distribution of the free parameters} \aT and \bT, given $T$. Note that we also need this distribution to infer the values of physical parameters from experiment by calculating their posterior probability.\footnote{Thus, we cannot use experimental constraints to say that we have empirical constraints on the prior right from the start. As the formula shows, we cannot turn empirical evidence $U$ into information about the value of the constant in our universe $p(\alpha|UTB)$ without a prior derived purely from the theory and theoretical background information $p( \alpha | T B)$.} For example, in one dimension,
\begin{equation} \label{eq:paramposterior}
p(\alpha|UTB) = \frac{p(U | \alpha T B) p( \alpha | T B)} { \int p(U | \alpha T B) p( \alpha | T B) \dd \alpha} ~.
\end{equation}

\section{Fine-Tuning as Physics Jargon} \label{S:FTjargon}

Philosophical discussions of the fine-tuning of the universe for life have often failed to recognize the context in which physicists have made their claims. `Fine-tuning', a metaphor that brings to mind a precisely-set analogue radio dial, is used as a technical term in physics. \citet{Donoghue2007} discusses the case of a theory in which a positive measurable quantity $x$ is calculated to be the sum of an unknown bare value ($x_0$) and an estimatable quantum correction ($x_q$). These quantities $x_0$ and $x_q$ are, according to the theory, unrelated. If, however, we discover that the measured value is much smaller than the quantum correction, then the theory is fine-tuned or \emph{unnatural} in a technical sense. To explain the data ($x$), we require that the following coincidence holds: $x_0 \approx -x_q$. However, this cancellation is unexpected and unexplained by the theory, suggesting that we should search for a theory that implies a deeper relationship between these quantities.

We can formalize this intuition using probability theory. Consider a simplified scenario in which a theory $T$ has a free parameter $\alpha$ that ranges from 0 to $R_\alpha$. The theory itself and our background mathematical knowledge $B$ give no reason to prefer any particular value of $\alpha$ in this range, so we represent our state of knowledge with a uniform probability distribution: $p(\alpha|T B) \dd \alpha = \dd \alpha / R_\alpha$. Suppose that the likelihood of the observed data is equal to 1 in a range of width $\Delta \alpha$, and zero outside.\footnote{The reader is invited to generalize this lesson to a narrow Gaussian likelihood, and to the case of multiple, coincidentally-related parameters, as was the case with $x$ above. Remember that the likelihood is normalized over \emph{data} $D$, not over free parameters. That is, it need not be the case that $\int p(D|\alpha T B) \dd \alpha = 1$.} Then,
\begin{equation} \label{eq:totalprobD}
p(D|TB) = \int_0^{R_\alpha} p(D | \alpha T B) p( \alpha | T B) \dd \alpha \approx \frac{\Delta \alpha}{R_\alpha} ~.
\end{equation}
If $\Delta \alpha \ll R$, then $p(D|TB) \ll 1$. That is what a physicist means by \emph{fine-tuned}. It is a special case of Bayesian theory testing: fine-tuned or unnatural theories have very small likelihoods, and the more fine-tuned, the smaller the likelihood. Now, a small likelihood is not a sufficient reason to discard a theory. Rather, it presents a explainable but as-yet-unexplained fact, a good reason to examine our background assumptions and/or search for an alternative theory, perhaps one whose free parameters are not as fine-tuned or in which the quantity $\alpha$ is not a free parameter at all.

In particular, the likelihood of the data given a theory with a free parameter becomes smaller when experiments measure the free parameter more precisely. This doesn't necessarily mean that the posterior probability of the theory becomes smaller. For example, if the relevant parameter is free in all known, viable physical theories, then measuring it more precisely won't make much of a difference. But if an alternative theory doesn't have the free parameter, then better measurement can increase its posterior probability. Suppose that in theory $T_1$, $\alpha$ is a free parameter which ranges between 0 and 1, while in theory $T_2$, $\alpha$ is constrained by a symmetry principle to be precisely 0.5. Suppose data $D$ implies $\alpha = 0.45 \pm 0.2$. (Continuing the simplification above, consider a step function that is one in the quoted range and zero outside.) This data is hardly decisive. But suppose the data $D_\textrm{new}$ from an improved experiment implies $\alpha = 0.50002 \pm 0.00005$. Supporters of $T_2$ are suitably thrilled, because while $p(D_\textrm{new}|T_2 B) = p(D|T_2 B) 
= 1$, $p(D_\textrm{new}|T_1 B) \ll p(D | T_1 B)$.

How are we to guard against theories that ``pre-cook" their parameters to match the data? As with the ad hoc theories discussed above, the prior is the key. Consider again theory $T_1$, and create a new theory $T'_1 = TA$ by adding the statement $A = $``$\alpha = 0.5 \pm 0.00005$". Then, the theory's likelihood is ideal $p(D_\textrm{new} | T'_1 B)$. However, the prior probability of the `theory' is now $p(T'_1 | B) = p(A|TB) p(T | B)$, and $p(A|T_1B) \approx p(D | T_1 B)$. The pre-cooked theory fares no better.

\section{Free Parameters and their Limits}
In the examples given above, we simply postulated the ranges of the free parameters $R_\alpha$. However, what if the range is infinite? Assigning a constant, normalized probability measure over an infinite range is impossible: there is no probability distribution $p$ such that a) $p(x)$ is constant, and b) $\int_0^\infty p(x) \dd x = 1$. So, how should we test a theory in such circumstances?

Under certain circumstances, a non-normalizable probability distribution can be tolerated. For example, if we are interested in using Equation \eqref{eq:paramposterior} to calculate the posterior probability distribution $p(\alpha | UTB)$ of the free parameter $\alpha$, then we may only need to assume that the prior probability distribution is constant near the peak of the likelihood $p(D | \alpha T B)$. In fact, many non-normalizable prior probability distributions (not just constant ones) can give sensible answers in such circumstances. However, this will not do when it comes to calculating the likelihood of the theory $p(D | T B)$ via Equation \eqref{eq:totalprobD}, since the constant value of the prior does not cancel out.

Infinities are well-known to produce apparent paradoxes in probability theory. In Chapter 15 of \citet{Jaynes2003}, a number of these paradoxes are carefully reviewed and discussed. Jaynes notes their common cause: passing from a well-defined probability problem (finite or convergent) to a limit –-- infinite magnitude, infinite set, zero measure, improper probability distribution function (PDF), or some other kind --– without specifying how the limit is approached. If one then forgets the limiting process, and asks a question whose answer depends on said process, contradictory results are inevitable but hardly surprising.\footnote{``It is not surprising that those who persist in trying to evaluate probabilities directly on infinite sets [and] trying to calculate probabilities conditional on propositions of probability zero, have before them an unlimited field of opportunities for scholarly looking research and publication --- without hope of any meaningful or useful results." (pg. 485)}

Jaynes's conclusion: ``based on some 40 years of mathematical efforts and experience with real problems –-- is that, at least in probability theory, an infinite set should be thought of only as the limit of a specific (i.e. unambiguously specified) sequence of finite sets." In particular, measure theory is to be used with caution, as it can easily disguise infinities. Further, ``in practice we will always have some kind of prior knowledge \ldots [that implies that the location and scale] parameters $(a,b)$ cannot vary over a truly infinite range.'' We can confine our attention to ``the range which \ldots expresses our prior ignorance" (pg. 396).

We intend here to take Jaynes's advice. But what could this extra knowledge be that defines the limiting process or constrains the free parameters to a finite range? Keep in mind that fundamental constants and initial conditions are defined by their theories; they have no ``theory-independent" existence.\footnote{The \emph{measurement} of a particular constant, stripped to its barest elements, is theory independent. For example, we can still measure the deflection of a torsion balance in the presence of a given mass. But only Newton's theory of gravity will tell us how to combine our measurements into the fundamental constant $G$.} Thus, in the context of testing a physical theory by calculating the likelihood of the data $p(D|TB)$, the only ``extra'' information that could constrain the parameters is \emph{the theory itself}. A physical theory, to be testable, must be sufficiently well-defined as to allow probabilities of data (likelihoods) to be calculated, at least in principle. Otherwise, the theory cannot tell us what data we should expect to observe, and so cannot connect with the physical universe. If the theory contains free parameters, then since the prior probability distribution of the free parameter is a necessary ingredient in calculating the likelihood of the data, the theory must justify a prior. In summary, a theory whose likelihoods are rendered undefined by untamed infinities simply fails to be testable. In essence, it fails to be a physical theory at all.

Note that we have not yet raised the fine-tuning of the universe for life. These are the conditions for a viable, testable physical theory.

How, then, do current theories of physics avoid problematic infinities? Consider Newton's theory of gravity ($N$), and its free parameter $G$. At first glance, it may appear that $G$ could be any positive real number, and no specific number is preferred, and hence Newtonian gravity fails to provide a normalizable prior probability distribution for its parameter $p(G|NB)$. However, $G$ has physical units: in SI units, m$^3$ kg$^{-1}$ s$^{-2}$. By changing our system of units, we can give $G$ any value we please. Thus, $p(G|NB)$ is ill-posed because $G$ is arbitrary. We need more physics to specify a system of units (such as the Planck units), or a specific experimental setup.

Suppose we are presented with the Cavendish experiment $C$: a rod of length $L$ with masses $m$ at each end is suspended from a thin wire, whose torsion creates a natural oscillation period $T$. Two masses $M$ are brought to a distance $r$ from the masses $m$, and the angular deviation of the rod $\theta$ is measured. The inferred value of $G$ is $2\pi L r^2 \theta / (M T^2)$. Knowing only the details of the setup $C$, but not the data $\theta$, is the possible range of $G$ infinite? No, because if $G$ were sufficiently large, the masses $m$ would crush the rod. Since we are taking $C$ as given [$p(G|CNB)$], the fact that $C$ is a stable experimental apparatus \emph{at all} places an upper limit on $G$, removing the problem of an infinite possible range for the free parameter.\footnote{It also places a lower limit on negative values for $G$, since strongly repulsive gravity would also destroy the rod. Note that this is not the usual Bayesian way of testing theories, as we would be required to calculate the prior $p(N|CB)$. This is almost ``cheating" --- we are smuggling data (the existence of the apparatus) into the calculation of the prior. At best, given that the apparatus, considered in isolation from the rest of the universe, relies only on non-gravitational physics, we might argue that $C$ is irrelevant and hence $p(N|CB) = p(N|B)$. The point of this example is that $p(G|NB)$ is not well-posed; this point is unaffected.}

The moral of the story is that, whether measuring constants and testing the theory itself, many physical theories cannot be considered in isolation. Our ``theory" $T$ must be able to describe not just the system of interest but also the measuring apparatus. For example, in calculating the likelihood of galaxy redshifts in cosmology, we not only need Einstein's General Relativity to describe expanding space, but also quantum theory to permit the interpretation of atomic emission spectra, from which redshifts are derived.

In modern physical theories, we can define our system of units using three fundamental parameters. A particularly useful system involves setting $G = c = \hbar = 1$, where $c$ is the speed of light and $\hbar$ is the reduced Planck constant. This defines \emph{Planck units}, in which the standard units of mass, length and time are $m_\textrm{Pl} = 2.17651 \times 10^{-8}$ kg, $1.616199 \times 10^{-35}$ m and $5.39106 \times 10^{-44}$ s, respectively. It then makes no sense to vary $(G,c,\hbar)$, to ponder what would happen if they were different, or to ask for their prior probabilities. The advantage is that, once units have been fixed, varying the other fundamental parameters (electron mass, quark masses, strengths of the fundamental forces etc.) makes a real difference to the universe, rather than simply making an essentially identical, scale-model of our universe.

\citet{Tegmark2006} list 26 parameters of the standard model of particle physics, and five parameters of the standard model of cosmology.\footnote{Cosmologists usually consider six parameters, adding the Hubble parameter $H_0$ to Tegmark et al's list. This parameter is not a constant, and essentially measures the age of the universe today. Its value is thus linked to our existence as observers, and so is not predicted by fundamental theory. If it is predicted at all, it is by anthropic arguments \citep{Lineweaver2007}.} Of these, one particle physics parameter and four cosmological parameters have units: the Quadratic Higgs coefficient (or, equivalently, the Higgs vacuum expectation value $v$), the cosmic dark energy density $\rho_\Lambda$, baryon mass per photon $\xi_b$, cold dark matter mass per photon $\xi_c$, and neutrino mass per photon $\xi_\nu$. In fact, the cosmological ``per photon'' quantities are the result of asymmetry in physics. For example, $\xi_b$ is derived from matter-antimatter symmetry: $\xi_b \sim m_\textrm{proton} ~ \eta$, where $\eta = (n_b - n_{\bar{b}}) / n_{\gamma}$ is the dimensionless Baryon asymmetry parameter. \citet{Tegmark2006} consider a model for dark matter in which $\xi_c$ is linked to the Peccei-Quinn (PQ) symmetry breaking scale ($f_a$). We thus will consider these parameters to be derived from more fundamental, dimensionless asymmetry parameters.

This leaves the dimensional parameters $v$ and $\rho_\Lambda$. Is there anything in the relevant physical theories that limits their range of possible values? Yes --- the Planck scale. Famously, we do not have a quantum theory of gravity. That is, we do not know how to describe gravity within a quantum framework. Naively combining quantum mechanics and general relativity, we can calculate that if a single particle were to have a mass equal to the Planck mass, then its black hole (Schwarzschild) radius would be larger than its quantum size (Compton length), and thus it would become its own black hole. The point is not that we think that this would actually happen, but rather that can't possibly trust our theories in this case. The Planck mass represents an upper boundary to any single-particle mass scale in our current theories. A lower boundary is provided by zero, since quantum field theory breaks down for negative masses; even if it didn't, a lower bound would be given by $-m_\textrm{Planck}$. Thus, the theory itself constricts the value of $v$ to the range $[0,m_\textrm{Planck})$, and $\rho_\Lambda$ to the range $[0,m_\textrm{Planck}^4)$ \citep{Wilson1979,Weinberg1989,Dine2015}. Outside of these ranges, our current theories cannot be trusted. Indeed, the very concepts that underlie our constants --- space, time, mass, energy --- may cease to be meaningful in the quantum gravity regime. This constraint comes from $T$ itself

What about dimensionless parameters? Some are phase angles, which have a finite range $[0, 2\pi)$. Others are coupling constants, which might vary over an infinite range. Two considerations can result in a normalizable prior distribution. Firstly, the masses of certain particles depend on the coupling constants, and increasing the constant beyond a certain value will send the mass beyond the Planck scale. For example, the mass of the electron is $m_e = v \Gamma_e /\sqrt{2}$, where $\Gamma_e$ is the dimensionless electron Yukawa coupling. If $\Gamma_e > \sqrt{2} m_\textrm{Planck} / v$, then the electron mass is greater than the Planck mass, and we have reached the border of our theory. Secondly, dimensionless parameters are expected to be of order unity. This is the idea behind the definition of \emph{naturalness} due to \citet{tHooft1980}:
\begin{quote}
a physical parameter or set of physical parameters is allowed to be very small [compared to unity] only if the replacement [setting it to zero] would increase the symmetry of the theory.
\end{quote}
\citet{BT86} offer the following justification for a preference for order-unity parameters in physics. When physicists use dimensional analysis to predict the form of an equation, we expect that the associated dimensionless constant is a combination of a few geometric factors (2, $\pi$, etc), with each contribution equally likely to be multiplied or divided. Thus, we consider dimensionless constants that are many orders of magnitude away from one to be \emph{unnatural} \citep[see, for example, the discussion in][]{Dine2015}. Perhaps simplicity also plays a role: a constant that is equal to unity is effectively no constant at all! Even if the possible range for a dimensionless constant is infinite, the principle of indifference does not force us to put a constant prior probability distribution over this range. We can debate the most appropriate form of the distribution, but there are plenty of functions that are non-zero over an infinite range and yet normalizable.

In short, there are normalization problems for a physical theory with free parameters that \emph{both} vary over an infinite range and are uniformly distributed. The standard models of particle physics and cosmology avoid these problems as follows. Dimensional parameters do not vary over an infinite range; they are bounded by the Planck scale. Dimensionless parameters might not vary over an infinite range, and common practice in the physical sciences assumes that parameters of order unity are more probable, so a uniform probability distribution is not forced upon us by the principle of indifference.

What about reparameterizations? Given any finite range of a free parameter, say $\alpha \in [a,b]$, we can write an equivalent theory in terms of a parameter $\gamma$ which varies over an infinite range. Here, we again rely on the theory to tell us which of these parameterizations is simpler. Often there are symmetries and invariances to guide our choice, for example the famous prior of \citet{Jeffreys1946}. Given that it cannot be the case that both $p(\alpha|TB) \dd \alpha$ and $p(\gamma|TB) \dd \gamma$ are constant, both cannot honestly represent our state of knowledge. We must decide, or more exactly, the theory itself must aid our determination of the prior. More than this cannot be said without considering the details of individual theories. But as \citet{Collins2009} has pointed out, by reparameterizing a theory we could make any prediction, even one as successful as QED's one in a billion prediction of the gyromagnetic moment of the electron, seem trivial.

\section{Fine-Tuning for Life}

We turn to the fine-tuning of the universe for life. We can follow the formalism of Section \ref{S:FTjargon}, applying it to this fact about our universe: it has developed and supports physical life forms ($L$). For our purposes, a typical dictionary definition will do: a living entity has the capacity to grow, metabolize, actively resist outside disturbance, and reproduce. A precise definition of life is not required, for the following reason. We would like to be able to place firm boundaries in parameter space between possible universes that would develop and support life and those that would not. However, this is not practically possible, as we do not know the sufficient conditions for abiogenesis. What we can do is consider a conservative outer boundary associated with sufficient conditions for lifelessness. For example, if the cosmological constant were negative, and its absolute value $10^{90}$ times smaller than the Planck scale (rather than $10^{120}$ in our universe), space would recollapse into a big crunch in one minute. This, it seems, is a sufficient condition for a lifeless, physical-observer-less universe. 

We consider the likelihood that a particular universe is life-permitting (or, in practice, not life-prohibiting), given the laws of nature as we know them. As above, we marginalize over the free parameters,
\begin{equation} \label{eq:totalprobL}
p(L|TB) = \int p(L | \aT \bT T B) ~ p( \aT \bT | T B) \dd \aT \dd \bT ~.
\end{equation}
The scientific literature on fine-tuning has identified life-permitting regions of the constants, outside of which life is seemingly physically impossible, that is, $p(L | \aT \bT T B)$ is extremely small. The most significant constants on which such constraints can be made are as follows. From particle physics: the Higgs vacuum expectation value (vev), the masses (or, equivalently, Yukawa parameters) of the electron, neutrinos (the sum of the three species), and up-, down- and strange-quark, the strong force coupling constant, and the fine-structure constant. From cosmology: the cosmological constant, the scalar fluctuation amplitude (“lumpiness”, $Q$), the number of spacetime dimensions, the baryonic and dark matter mass-to-photon ratios, and the initial entropy of the universe. These cases are discussed in \citet{Hogan2000}, \citet{Tegmark2006}, \citet{Barnes2012} and \citet{Schellekens2013}.

In particular, for the Higgs vev ($v$) and the cosmological constant ($\rho_\Lambda$), the Planck scale is not merely the maximum possible value of the parameter. Quantum corrections contribute terms of order $m_\ro{Planck}^2$ and $m_\ro{Planck}^4$ to $v^2$ and $\rho_\Lambda$ respectively, meaning that the Planck scale is the \emph{natural} scale for these parameters \citep{Dine2015}. The smallness of these parameters with respect to the Planck scale is known as the hierarchy problem and the cosmological constant problem respectively. Their life-permitting limits are as follows. If $v^2/m_\ro{Planck}^2 \lesssim 6 \ten{-35}$, then hydrogen is unstable to electron capture; if $v^2/m_\ro{Planck}^2 \gtrsim 10^{-33}$ then no nuclei are bound and the periodic table is erased. If $\rho_\Lambda / \rho_\ro{Planck} \lesssim - 10^{-90}$, the universe would recollapse after 1 second; if $\rho_\Lambda / \rho_\ro{Planck} \gtrsim 10^{-110}$, then no structure whatsoever would form in the universe. These limits come from \citet{Hogan2000}, \citet{Tegmark2006}, \citet{Barnes2012}, \citet{Schellekens2013}, and references therein.

In light of the naturalness of the Planck scale, our ignorance of these parameters can be honestly represented by prior probability distributions that are uniform in $v^2/m_\ro{Planck}^2$ and $\rho_\Lambda / \rho_\ro{Planck}$, resulting in probabilities $\sim 10^{-33}$ and $10^{-90}$ respectively. Indeed, a uniform prior seems to be conservative. We could argue for a prior that peaks at the Planck scale, which would decrease these probabilities even further.

These probabilities are surprisingly small in the following sense. $L$ could have been a generic fact about universes which obey the laws of nature as we know them. That is, it could have been the case that some form of life would exist for a wide range of values of the free parameters. The smallness of the likelihood of $L$ is not the result of considering a highly specified, precise outcome. For example, the likelihood of experimental data can depend on the precision of our measurements. The probability of a radioactive nucleus decaying in a given 1 second interval is smaller than the probability in a given 1 minute inverval. By contrast, the likelihood of $L$ does not depend on the precision of our measurements.\footnote{Note that the degree of fine-tuning is not the degree of accuracy of various order-of-magnitude models used in the early fine-tuning literature, contra \citet{Klee2002}.}

The smallness of these likelihoods, as noted above, proves nothing by itself. But very small probabilities should make us suspicious --- perhaps something unlikely happened, or perhaps one of the assumptions that went into the probability calculation needs to be challenged.

\section{Responses to Critics}
\subsection{McGrew, McGrew \& Vestrup, and Colyvan, Garfield \& Priest}

MMV consider the fine-tuning argument, contending that we cannot conclude anything from the narrow life-permitting ranges of the fundamental parameters because ``the Euclidean measure function \ldots is not normalizable. If we assume every value of every variable to be as likely as every other --- more precisely, if we assume that for each variable, every small interval of radius $e$ on $R$ has the same measure as every other --- there is no way to `add up' the regions of $R^K_+$ so as to make them sum to one. If they have any sum, it is infinite."

Similarly, CGP argue that ``the probability of finding the constant in question in \emph{any} finite interval is zero. This makes a mockery of the claim that \emph{the class of life-permitting universes, in particular}, is improbable."

MMV and CGP's concerns are perfectly legitimate, but are not specifically a problem with fine-tuning. They have identified the conditions under which a physical theory, scuppered by infinities, fails to produce likelihoods of data --- any data. Consider Equations \eqref{eq:totalprob} and \eqref{eq:totalprobL}: if $p( \aT \bT | T B)$ is undefined, we can't calculate $p(L|TB)$ because we can't calculate $p(D | TB)$ for any data $D$. If the fundamental theories of modern physics cannot \emph{in principle} justify a normalizable prior distribution over their free parameters, then we have bigger problems than fine-tuning. We don't have a testable theory at all. We can't derive predictions, can't model data, and must go back to square one.

We have seen above how modern theories avoid the problems raised by MMV and CGP. The set of possible values of a constant is dictated by the theory. If the model (or combination of models, such as gravity plus quantum mechanics) fails to be mathematically consistent beyond a certain value of the constant, then this limits the constant's possible range. The constant only lives within the theory, so where the theory ceases to be coherent, the constant ceases to be possible. In particular, mass-energy (including energy density) parameters are bounded between zero and the Planck scale. With dimensionless parameters, we should not assume that every value is as likely as every other, because `order-unity' ($\sim 1$) values are more likely.

CGP consider the idea that ``the laws of physics themselves set limits on the values certain constants can take'' as a possible solution to the problem of non-normalizability. They reply that ``without some independent argument for the shape of the distribution in question, this version of the argument simply begs the question."

On the contrary, \emph{some} resolution to this problem is necessary for any law of physics to be tested using Bayesian probabilities. Without a prior distribution for the free parameters, no likelihood can be calculated and no predictions made. Moreover, the shape of the distribution need only honestly reflect our ignorance of the parameter in the absence of experimental data. We can debate the shape that fulfills this requirement; the problem of prior probabilities in Bayesian data analysis is an open research question, with ``catalogues" of distributions available to the discerning scientist \citep{Yang1997}. We have argued above that, for ``natural'' parameters (in the physicist's sense of the word), a uniform distribution over the finite range dictated by the laws themselves is reasonable and indeed conservative. But for such a distribution not to exist at all would hamstring the entire Bayesian approach to testing physical theories.

\subsection{Halvorson}

Many of the criticisms of \citet{Halvorson2014} are relevant only to the fine-tuning argument for God, which is beyond our purview here. We comment only on the important distinction between credences and chances in testing physical theories in general and in the problem of fine-tuning in particular. Here, we use the following definition of David Lewis's Principal Principle: ``a person Ms degree of belief (at t) that A conditional on the proposition that the chance of A (at t) is x should be equal to x; with the qualification that at t she has no information about A that is inadmissible. Information about A is inadmissible if it is information about A over and above information about A's chance.'' \citep{Loewer2004}.

Halvorson considers an urn containing with 1 yellow ball and 99 purple balls. Because the balls are otherwise identical, ``any two balls are equally likely to be drawn''. Thus, ``according to the principal principle, the rational credence for drawing a yellow ball, given [background information], is 0.01." This probability is unaffected by any additional hypothesis regarding how the balls were placed in the urn. Similarly, says Halvorson, the probability of the universe and its constants is set by the laws of nature, and is unaffected by any deeper story about how the universe came to be.

Note that Halvorson's view allows for a positive scientific case to be made for the multiverse. If a cosmologist proposes a theory on which a multiverse is almost inevitable, then the theory could assign an almost unity likelihood to our observation of a life-permitting universe. Thus, that theory would be confirmed over a theory that implied a single, fine-tuned universe. But note carefully what the cosmologist cannot do: they cannot take a certain physical theory, find a set of initial conditions that will produce a multiverse, and then propose that such initial conditions should be privileged in some way. This, for Halvorson, is inadmissible: the theory sets the probability of initial conditions, and no further assumption (theism, multiverse or whatever) can change this assignment. It should be noted that advocates of the \emph{past hypothesis} do exactly what Halvorson prohibits: ``the distribution of probabilities over all of the possible exact initial microconditions of the world is uniform \ldots over those possible microconditions of the universe which are compatible with the initial [very low-entropy] macrocondition \ldots, and zero elsewhere.'' \citep{Albert2015}

My central concern here is the claim that chances of initial conditions are set by the theory. This is mistaken, and for the same reason that the claim about the urn is mistaken. It does not follow from the fact that the ball are otherwise identical that any two balls are equally likely to be drawn. That assumption is \emph{additional} to the setup specified by Halvorson's background information (1 yellow, 99 purple, otherwise identical). Putting balls in an urn, even near-identical ones, does not specify the mechanism by which a ball is selected, and so does not specify any chances at all.

If we remove the assumption that ``any two balls are equally likely to be drawn'', we do not have any chances, but we can still reason about credences as follows. Either the ball will be drawn by a process that is indifferent $(I)$ to the colour of the ball, or it will not $(\bar{I})$. On $I$, the chances of all balls are equal, and so by the principle of indifference, $p(Y|I ~ B) = 0.01$, where $B$ is the relevant background information about the urn and its contents. On $\bar{I}$, however, we have a process that is trying to select one colour of ball. We do not know which colour is preferred, so we say that the probability of a yellow preference ($P_Y$) is equal to the probability of a blue preference ($P_B$), so that $\bar{I} = P_Y + P_B$. Then, by the law of total probability:
\begin{equation} \label{eq:probyellow}
p(Y | B) = p(Y | IB)~ p(I|B) + p(Y|P_Y B) ~ p(P_Y | B) + p(Y | P_B B) ~ p(P_B | B) ~,
\end{equation}
where, by assumption, $p(I|B) = 1/2$ and $p(P_Y | B) = p(P_B | B) = 1/4$. The remaining unknowns are $p(Y|P_Y B)$ and $p(Y|P_B B)$, the probability of someone biased towards yellow successfully drawing yellow, and the probability of someone biased towards blue accidentally drawing yellow. These depend on the urn --- if locating and selecting the yellow ball is made near-impossible by (say) a deep, dark interior and a narrow neck, then even a biased selector will do no better than random: $p(Y|P_Y B) = p(Y|P_B B) \approx 0.01$, in which case $p(Y | B) = 0.01$. However, if the urn is transparent and the balls easily manoeuvred, then $p(Y|P_Y B) \approx 1$ and $p(Y|P_B B) \approx 0$, in which case $p(Y | B) \approx 0.25$.

This is the Bayesian approach: we model our state of knowledge, marginalizing over our assumptions. What we do not do is invoke the principal principle, set $p(Y | B) = 0.01$, and use this as a constraint on the probability of the various scenarios for how the ball was drawn. Further, the existence of a measure $\mu$ over the set of possible draws from the urn which respects a certain symmetry (in this case, permutation independently of colour) implies neither chances nor credences. The assumption of equal chances is \emph{additional} to the facts about the urn, including the measure.

We apply this to the laws of nature. A measure over the set of initial conditions of a theory, even one that follows naturally from the theory itself in some sense (respecting symmetry, for example), does not imply the assignment of chances to those initial conditions. Indeed, such a claim would be highly problematic. Are we to think that all theories imply an actually existing ensemble of universes, with properties distributed in accordance with the measure over initial conditions? That would add significant ontological baggage to our laws. But surely this is mistaken. Suppose I set up a computer simulation of the universe. I choose the laws that my simulation will follow, and those laws may motivate a measure over their space of initial conditions. It does not follow that I am obliged to select initial conditions according to a chancy process that respects that measure. Furthermore, by definition, no physical mechanism sets initial conditions, especially of the whole universe. So it is difficult to understand how this could be understood as a ``single-case'' probability or chance or propensity.

The connection between the measure and credences is even less tight. It is sometimes supposed that fine-tuning arguments assume that for our universe to be improbable, one must postulate that the universe is selected at random by some stochastic physical mechanism. Our ignorance of that mechanism, then, would seem to make fine-tuning a purely speculative exercise. But Bayesian model selection does not rely solely on physical chances. Indeed, testing physical theories cannot use chances alone if it wants to calculate the probability \emph{of} a theory. Outcomes, or physical events, can be chancy, but theories are not.

\section{Conclusion}
When faced with the question ``why is the universe the way it is?", we might want to consider the related question ``what other ways could the universe have been?". To approach this question systematically, we can take the deepest known laws that describe how the physical universe works and look for ways to vary them. A particularly tractable way is to vary their free parameters. The set of solutions to those laws, usually represented by the set of possible initial conditions, is precisely the set of scenarios that the law deems ``physically possible''. And the constants that appear in the laws themselves have long attracted the attention of theoretical physicists as being in need of deeper explanation.

It is an interesting fact, then, that this search for other ways that the universe could have been has overwhelmingly found lifelessness. This lifelessness is surprising in the way that any fine-tuned parameter in physics is surprising: it is improbable. By the standards of Bayesian model selection, an explanation of this fact should be sought.

\section*{Acknowledgments}
Supported by a grant from the John Templeton Foundation. This publication was made possible through the support of a grant from the John Templeton Foundation. The opinions expressed in this publication are those of the author and do not necessarily reflect the views of the John Templeton Foundation.

\end{document}